\documentclass[%
 aip,
 jmp,%
 amsmath,amssymb,
 reprint,%
]{revtex4-2}

\usepackage{graphicx}
\usepackage{dcolumn}
\usepackage{bm}
\usepackage{comment}
\usepackage{xcolor}

\begin{document}

\preprint{AIP/123-QED}

\title{Pressure-Strain Interaction: I. On Compression, Deformation, and Implications For Pi-D}

\author{Paul A. Cassak}
\email{Paul.Cassak@mail.wvu.edu}
\author{M. Hasan Barbhuiya}
\affiliation{Department of Physics and Astronomy and the Center for KINETIC Plasma Physics, \\
West Virginia University, Morgantown, WV 26506, USA
}

\date{\today}

\begin{abstract}
The pressure-strain interaction describes the rate per unit volume that energy is converted between bulk flow and thermal energy in neutral fluids or plasmas. The term has been written as a sum of the pressure dilatation and the collisionless analogue of viscous heating referred to as ${\rm Pi-D}$, which isolates the power density due to compressible and incompressible effects, respectively. It has been shown that ${\rm Pi-D}$ can be negative, which makes its identification as collisionless viscous heating troubling.  We argue that an alternate decomposition of pressure-strain interaction can be useful for interpreting the underlying physics. Since ${\rm Pi-D}$ contains both normal deformation and shear deformation, we propose grouping the normal deformation with the pressure dilatation to describe the power density due to converging/diverging flows, with the balance describing the power density purely due to shear deformation. 
We then develop a kinetic theory interpretation of compression, normal deformation, and shear deformation. We use the results to determine the physical mechanisms that can make ${\rm Pi-D}$ negative. We argue that both decompositions can be useful for the study of energy conversion in weakly collisional or collisionless fluids and plasmas, and implications are discussed.
\end{abstract}

\keywords{Energy conversion, dissipation, magnetic reconnection, plasma turbulence, collisionless shocks}

\maketitle

\section{Introduction}
\label{sec:intro}

Weakly collisional plasmas are important in many settings, from heliophysics to planetary magnetospheres to astrophysics \cite{Howes17}.  A host of plasma phenomena take place in such settings, including magnetic reconnection, plasma turbulence, and collisionless shocks. The dearth of collisions in many settings of interest implies that these plasmas can be far from local thermodynamic equilibrium (LTE). In the study of these physical phenomena, one of the forefront research questions is how energy is converted during each process, especially when non-LTE effects greatly affect the dynamics at the micro-, meso-, and even the macro-scale \cite{Matthaeus20}.

A quantity contributing to non-LTE energy conversion that has received intense scrutiny over the last few years is the pressure-strain interaction, written as $-({\bf P} \cdot \boldsymbol{\nabla}) \cdot {\bf u}$, where ${\bf P}$ is the pressure tensor of a species of a fluid or plasma and ${\bf u}$ is its bulk flow velocity \cite{del_sarto_pressure_2016,Yang17,yang_PRE_2017,del_sarto_pressure_2018}. In terms of the phase space density $f$ (the number of particles per unit position space volume and velocity space volume), the bulk flow velocity is ${\bf u} = (1/n) \int d^3v {\bf v} f$, where ${\bf v}$ is the velocity space coordinate, $n=\int d^3v f$ is the number density, and the integrals are over all velocity space, and the elements of the pressure tensor ${\bf P}$ are (classically and non-relativistically) $P_{jk} = m \int v_{j}^\prime v_{k}^\prime f d^3v$, where $j,k$ are indices for the spatial dimensions, $m$ is the constituent particle mass, and ${\bf v}^{\prime} = {\bf v} - {\bf u}$ is the peculiar (random) velocity.  

To see why pressure-strain interaction is important for energy conversion, consider the thermal (internal) energy density ${\cal E}_{th} = 3 {\cal P} / 2 = \int (\frac{1}{2} m v^{\prime 2}) f d^3v$, where ${\cal P} = (1/3) {\rm tr} ({\bf P}) = (1/3) P_{jj}$ is the effective pressure, using the Einstein summation convention for repeated indices here and throughout. Its time evolution is described by \cite{Braginskii65}
\begin{equation}
    \frac{\partial {\cal E}_{th}}{\partial t} + \boldsymbol{\nabla} \cdot ({\cal E}_{th} {\bf u}) = - ({\bf P} \cdot \boldsymbol{\nabla}) \cdot {\bf u} - \boldsymbol{\nabla} \cdot {\bf q} + \dot{Q}_{{\rm visc,coll}}, \label{eq:ethermaleqn}
\end{equation}
where ${\bf q} = \int (1/2) m v^{\prime 2} {\bf v}^\prime f d^3v$ is the vector heat flux density and $\dot{Q}_{{\rm visc,coll}}$ is the volumetric viscous heating rate via collisions, where we use the word viscous regardless of the functional form of the collisional heating.  The time evolution of the bulk kinetic energy density ${\cal E}_{k} = (1/2) m n u^2$ is given by \cite{Braginskii65}
\begin{equation}
\frac{\partial {\cal E}_k}{\partial t} + \boldsymbol{\nabla} \cdot \left({\bf u} {\cal E}_k + {\bf u} \cdot {\bf P}\right)  = \left({\bf P} \cdot \boldsymbol{\nabla}\right) \cdot {\bf u}  + n {\bf u} \cdot {\bf F} + {\bf R}_{{\rm coll}}, \label{eq:ekineticeqn}
\end{equation}
where ${\bf F}$ is the net body force and ${\bf R}_{{\rm coll}}$ is the inter-species collisional drag force power density. The pressure-strain interaction $-({\bf P} \cdot \boldsymbol{\nabla}) \cdot {\bf u}$ arises in each equation with opposite signs, so it describes the rate per unit volume that energy is converted between bulk flow and thermal. The minus sign is included in the definition so a positive value describes a contribution towards increasing thermal energy density.  While these equations have been known for years, a watershed moment came recently when it was emphasized that the pressure-strain interaction is key to describing changes in thermal energy in plasmas \cite{del_sarto_pressure_2016,Yang17,yang_PRE_2017,del_sarto_pressure_2018}.

The pressure-strain interaction can be simplified by defining the strain rate tensor $\boldsymbol{\nabla} {\bf u}$, which can be decomposed \cite{Batchelor67,del_sarto_pressure_2018} as $\boldsymbol{\nabla} {\bf u} = {\bf S} + \boldsymbol{\Omega}$ into a symmetric (irrotational) strain rate tensor ${\bf S}$ with elements $S_{jk} = (1/2) (\partial u_k / \partial r_j + \partial u_j / \partial r_k)$ and an anti-symmetric strain rate tensor $\boldsymbol{\Omega}$ with elements $\Omega_{jk} = (1/2) (\partial u_k / \partial r_j - \partial u_j / \partial r_k)$. A flow with non-zero ${\bf S}$ but zero $\boldsymbol{\Omega}$ has ``pure straining motion'' \cite{Batchelor67}; it strains a fluid element without rotating it.  In contrast, a flow with non-zero $\boldsymbol{\Omega}$ but zero ${\bf S}$ is ``rigid body rotation'' \cite{Batchelor67}, which rotates a fluid element without changing its shape.  A further decomposition of ${\bf S}$ was introduced \cite{Batchelor67,del_sarto_pressure_2016,Yang17,del_sarto_pressure_2018} by writing ${\bf S} = (1/3) {\bf I} (\boldsymbol{\nabla} \cdot {\bf u}) + \boldsymbol{{\cal D}}$, {\it i.e.,}
\begin{equation}
    S_{jk} = \frac{1}{3} \delta_{jk} (\nabla \cdot {\bf u}) + {\cal D}_{jk}, \label{eq:straindecomp}
\end{equation}
where ${\bf I}$ is the identity tensor, $\delta_{jk}$ is the Kroenecker delta and $\boldsymbol{{\cal D}}$ is the traceless strain rate tensor with elements
\begin{equation}
  {\cal D}_{jk} = \frac{1}{2} \left( \frac{\partial u_{j}}{\partial r_k} + \frac{\partial u_{k}}{\partial r_j} \right) - \frac{1}{3} \delta_{jk} (\boldsymbol{\nabla} \cdot {\bf u}). \label{eq:ddef}
\end{equation}
Physically, $(1/3) {\bf I} (\nabla \cdot {\bf u})$ describes compression/expansion, while $\boldsymbol{{\cal D}}$ describes the incompressible deformation of a fluid element \cite{Batchelor67,del_sarto_pressure_2018}, which is a volume preserving change of shape of the fluid element.

The pressure-strain interaction is then written in a number of equivalent ways.  In terms of the strain rate tensor, $-({\bf P} \cdot \boldsymbol{\nabla}) \cdot {\bf u} = - {\bf P} : \boldsymbol{\nabla} {\bf u} = -P_{jk} (\partial u_k / \partial r_j)$.  Using $\boldsymbol{\nabla} {\bf u} = {\bf S} + \boldsymbol{\Omega}$, it is immediately found that ${\bf P} : \boldsymbol{\Omega} = 0$ since ${\bf P}$ is symmetric under interchange of indices, so rigid body rotation does not contribute to pressure-strain interaction \cite{del_sarto_pressure_2016}. Consequently, $-({\bf P} \cdot \boldsymbol{\nabla}) \cdot {\bf u} = -{\bf P} : {\bf S} = - P_{jk} S_{jk}$, {\it i.e.,} pressure-strain interaction only has contributions from the pure straining motion portion. Further, one decomposes the pressure tensor as
\begin{equation}
  {\bf P} = {\cal P} {\bf I} + \boldsymbol{\Pi},
  \label{eq:pdecomppi}
\end{equation}
where $\boldsymbol{\Pi}$ is the deviatoric pressure tensor that describes the non-isotropic part of the pressure tensor. While the diagonal elements of ${\bf P}$ must be non-negative, all elements of $\boldsymbol{\Pi}$ can be either positive or negative. Using the decomposition of ${\bf S}$ in Eq.~(\ref{eq:straindecomp}) with $\boldsymbol{{\cal D}}$ defined in Eq.~(\ref{eq:ddef}) and the pressure decomposed in Eq.~(\ref{eq:pdecomppi}), one finds
\begin{equation}
  -({\bf P} \cdot \boldsymbol{\nabla}) \cdot {\bf u} = - {\cal P} (\boldsymbol{\nabla} \cdot {\bf u}) - \Pi_{jk} {\cal
    D}_{jk}, \label{eq:pdelu}
\end{equation}
where the cross-terms vanish because $\boldsymbol{\Pi}$ and $\boldsymbol{{\cal D}}$ are both traceless.  The benefit of this decomposition is that the first term (including the minus sign), called pressure dilatation, describes the power density of heating due to bulk compression ($\boldsymbol{\nabla} \cdot {\bf u} < 0$) or cooling due to bulk expansion ($\boldsymbol{\nabla} \cdot {\bf u} > 0$). The second term (including the minus sign) has been called \cite{Yang17} ${\rm Pi-D}$, which is the power density due to incompressible deformation  \cite{del_sarto_pressure_2018}.  ${\rm Pi-D}$ was also called ``collisionless viscosity'' because it is analogous in form to collisional viscous heating \cite{yang_PRE_2017}.

Much has been learned about the pressure-strain interaction and ${\rm Pi-D}$ in the context of plasma physics. The pressure-strain interaction was studied in strongly magnetized plasmas \cite{Hazeltine13}, including the recognition that the gyro-viscous contribution to the pressure-strain interaction vanishes identically. The fluid description of the contributions to the pressure-strain interaction was studied for the case with zero heat flux density \cite{del_sarto_pressure_2016,del_sarto_pressure_2018}. It was shown \cite{Yang17} that for a periodic or closed domain in a purely collisionless system, the volume average of Eq.~(\ref{eq:ethermaleqn}) implies that $<-({\bf P} \cdot \boldsymbol{\nabla}) \cdot {\bf u}>$ is the only term that can change the total thermal energy $E_{th} = \int d^3r {\cal E}_{th}$ of the system, where angular brackets denote a volume average.  Interestingly, the same study showed in simulations of plasma turbulence that ${\rm Pi-D}$ could be locally positive or negative (since elements of both $\boldsymbol{\Pi}$ and $\boldsymbol{{\cal D}}$ can be positive or negative).

Numerous studies have since investigated the pressure-strain interaction and ${\rm Pi-D}$ using numerical simulations. ${\rm Pi-D}$ is stronger in coherent structures (current sheets) than in the bulk in plasma turbulence \cite{yang_PRE_2017}. The pressure-strain interaction was highest in regions with current sheets and high vorticity \cite{Pezzi19}. ${\rm Pi-D}$ was found to successfully identify regions of strong energy conversion in dipolarization fronts \cite{Sitnov18,song_forcebalance_2020}. The pressure-strain interaction dominates other energy conversion metrics at small length scales \cite{yang_scale_2019}, and was shown to account for the net temperature increase in simulations of turbulence \cite{Pezzi19,Yang_2022_ApJ}.  A recent study compared the pressure-strain interaction during reconnection and turbulence, finding that pressure dilatation at current sheets was more important in turbulence than in reconnection \cite{Pezzi21}. ${\rm Pi-D}$ increases with plasma beta for ions, but the dependence is weak for electrons \cite{Parashar18}.  In island coalescence, pressure-strain interaction does not depend strongly on electron mass or system size \cite{Du18}. It was suggested that pressure-strain interaction contributes to the break in the turbulent spectrum at ion \cite{Hellinger22} and electron \cite{Arro21} scales, and therefore is a critical piece of the termination of the turbulent cascade \cite{Matthaeus20}. Importantly, ${\rm Pi-D}$ and the heat flux divergence have similar contributions in turbulence \cite{Du20}, and the heat flux divergence can oppose the pressure-strain interaction \cite{Fadanelli21}.

The pressure-strain interaction, including ${\rm Pi-D}$, has also been studied observationally, facilitated greatly by the high resolution measurements afforded by the Magnetospheric Multiscale (MMS) mission \cite{Burch16}.  In the turbulent magnetosheath, it was found that pressure-dilatation contributed more to the pressure-strain interaction than ${\rm Pi-D}$ \cite{Chasapis18}, as would later be seen in simulations \cite{Pezzi21}.  A statistical study of ${\rm Pi-D}$ in the turbulent magnetosheath found that it is spatially concentrated near current sheets as in the simulations, but is small within current sheets \cite{Bandyopadhyay20}, as would also later be reported in simulations \cite{Pezzi21}.  A study of magnetopause reconnection found that electrons were heated at a faster rate than ions and pressure-dilatation dominated ${\rm Pi-D}$ \cite{bandyopadhyay_energy_2021}.  The same study measured negative ${\rm Pi-D}$.  In a statistical study of reconnection diffusion regions, it was common to see a negative ${\rm Pi-D}$, and the pressure-strain interaction was positive in only about half of the events \cite{zhou_measurements_2021}.  They also found that the gyrotropic portion of ${\rm Pi-D}$ was more important than the non-gyrotropic part.  In a study of 50 turbulent magnetosheath events, both positive and negative intervals were found for both pressure dilatation and ${\rm Pi-D}$ \cite{Wang21}. A statistical study of 122 dipolarization fronts suggested that ${\rm Pi-D}$ is not a significant contributor to energy conversion \cite{Zhong19}.

Despite great advances in our knowledge about pressure-strain interaction in general, and ${\rm Pi-D}$ in particular, there are a number of puzzling aspects of its interpretation, especially ${\rm Pi-D}$. For example, it is not understood how ${\rm Pi-D}$ is effectively a collisionless viscosity but can be negative.
This study is the first in a three-part series on pressure-strain interaction. Here, we point out that the strain rate tensor contains both normal deformation and shear deformation, as is well known in continuum mechanics, which therefore implies that ${\rm Pi-D}$ contains power density due to both effects. This grouping of terms complicates the interpretation of ${\rm Pi-D}$ because it mixes stresses from normal flows and sheared flows. Because pressure dilatation is also associated with normal flows, we suggest an alternate decomposition of the pressure-strain interaction that groups the normal deformation with the pressure dilatation instead of shear deformation. This separates the effects of converging/diverging flow from shear strain. We calculate the terms in this alternate decomposition analytically. We then develop a physical interpretation of the compression and normal and shear deformation using kinetic theory. This allows us to determine the physical mechanisms that can make ${\rm Pi-D}$ negative, thereby clarifying how to interpret such measurements. In the second study \cite{Cassak_PiD2_2022} (``Paper II''), we write the pressure-strain interaction in magnetic field-aligned coordinates, which further elucidates the physical contributions to the pressure-strain interaction in a magnetized plasma.  In the third study \cite{Barbhuiya_PiD3_2022} (``Paper III''), we display the pressure-strain interaction and its Cartesian and magnetic field-aligned decompositions in simulations of reconnection.  We determine the physical causes for the pressure-strain interaction during reconnection.

The layout of this manuscript is as follows.  
An alternate decomposition of the pressure-strain interaction is derived in Sec.~\ref{sec:piddecomposition}.  We then provide a kinetic theory interpretation of the pressure-strain interaction contributions in Sec.~\ref{sec:physicalinterp}, and discuss the causes of ${\rm Pi-D}$ and the normal deformation being negative using kinetic theory in Sec.~\ref{sec:pidsign}. A discussion and conclusions are in Sec.~\ref{sec:discuss}. 


\section{An Alternate Decomposition of Pressure-Strain Interaction}
\label{sec:piddecomposition}

From the expression in Eq.~(\ref{eq:ddef}), we note the important general property, well known in continuum mechanics, that the diagonal elements of $\boldsymbol{{\cal D}}$ are associated with normal deformation while the off-diagonal elements are associated with shear deformation. To picture this, consider a cubic fluid element.  Normal deformation of the fluid element results from flow parallel to the normal to the edges of the fluid element that vary, while shear deformation results from flow in the plane of the edges of the fluid element that vary. Thus, we decompose $\boldsymbol{{\cal D}}$ into a normal deformation tensor $\boldsymbol{{\cal D}}_{{\rm normal}}$ and a shear deformation tensor $\boldsymbol{{\cal D}}_{{\rm shear}}$, so that 
\begin{equation}
\boldsymbol{{\cal D}} = \boldsymbol{{\cal D}}_{{\rm normal}} + \boldsymbol{{\cal D}}_{{\rm shear}}.
\end{equation}
Here, ${\cal D}_{{\rm normal},jk} = [(\partial u_j / \partial r_j) - (1/3) (\nabla \cdot {\bf u})]\delta_{jk}$ (with no sum on $j$) has the same diagonal elements as $\boldsymbol{{\cal D}}$ with its off-diagonal elements equal to zero and isolates normal deformation.  Similarly, ${\cal D}_{{\rm shear},jk} = (1/2)(\partial u_j / \partial r_k+\partial u_k / \partial r_j)$ for $j \neq k$ and ${\cal D}_{{\rm shear},jj} = 0$ (no sum on $j$) has its diagonal elements equal to zero and its off-diagonal elements equal to those of $\boldsymbol{{\cal D}}$, which isolates shear deformation. (A related decomposition was discussed in Refs.~\cite{Du18,zhou_measurements_2021}, but we do not make any assumptions about gyrotropy.)

In terms of this decomposition of $\boldsymbol{{\cal D}}$, we write ${\rm Pi-D}$ as the sum of two terms,
\begin{equation}
{\rm Pi-D} = {\rm Pi-D}_{{\rm normal}} + {\rm Pi-D}_{{\rm shear}},
\end{equation}
where ${\rm Pi-D}_{{\rm normal}} = - \boldsymbol{\Pi} : \boldsymbol{{\cal D}}_{{\rm normal}}$ and ${\rm Pi-D}_{{\rm shear}} = -\boldsymbol{\Pi} : \boldsymbol{{\cal D}}_{{\rm shear}}$. In Cartesian coordinates, a brief calculation reveals that these are
\begin{subequations}
\begin{eqnarray}
{\rm Pi-D}_{{\rm normal}} & = & -(\Pi_{xx} {\cal D}_{xx} + \Pi_{yy} {\cal D}_{yy} + \Pi_{zz} {\cal D}_{zz}) \nonumber \\ & = & -\left(\Pi_{xx} \frac{\partial u_x}{\partial x} + \Pi_{yy} \frac{\partial u_y}{\partial y} + \Pi_{zz} \frac{\partial u_z}{\partial z}\right), \label{eq:piddeform} \\
{\rm Pi-D}_{{\rm shear}} & = &  -(2 \Pi_{xy} {\cal D}_{xy} + 2 \Pi_{xz} {\cal D}_{xz} + 2\Pi_{yz} {\cal D}_{yz}) \nonumber \\ & = & -\left[P_{xy} \left( \frac{\partial u_x}{\partial y} + \frac{\partial u_y}{\partial x} \right) + P_{xz} \left( \frac{\partial u_x}{\partial z} + \frac{\partial u_z}{\partial x} \right) + P_{yz} \left( \frac{\partial u_y}{\partial z} + \frac{\partial u_z}{\partial y} \right) \right]. \label{eq:psincompress}
\end{eqnarray}
\end{subequations}
The terms separate the contributions due to normal deformation and shear deformation, respectively.

Mirroring the decomposition of ${\rm Pi-D}$, we revisit the pressure-strain interaction, which describes the full rate of conversion between bulk flow and thermal energy density.  Following Eq.~(\ref{eq:straindecomp}), we decompose the symmetric strain rate tensor ${\bf S}$ as
\begin{equation}
{\bf S} = \frac{1}{3} {\bf I} (\boldsymbol{\nabla \cdot {\bf u}}) + \boldsymbol{{\cal D}}_{{\rm normal}} + \boldsymbol{{\cal D}}_{{\rm shear}},
\end{equation}
Then, the pressure-strain interaction is decomposed into three pieces,
\begin{equation}
-({\bf P} \cdot \boldsymbol{\nabla}) \cdot {\bf u} 
= -{\cal P} (\boldsymbol{\nabla \cdot {\bf u}}) + {\rm Pi-D}_{{\rm normal}} + {\rm Pi-D}_{{\rm shear}}.
\end{equation}
These three terms isolate the power density due to dilatation, normal deformation, and shear deformation, respectively. 
A key point is that the normal deformation only depends on diagonal elements of $\boldsymbol{{\cal D}}$, {\it i.e.,} on converging/diverging flow, as seen in Eq.~(\ref{eq:piddeform}).  We thus argue that it may be more natural for the normal deformation to be combined with the pressure dilatation, which also only depends on the diagonal elements of $\boldsymbol{{\cal D}}$, than with ${\rm Pi-D}_{{\rm shear}}$.  We therefore introduce the quantity ${\rm PDU}$ as
\begin{subequations}
\begin{eqnarray}
{\rm PDU} & = & -{\cal P} (\boldsymbol{\nabla} \cdot {\bf u}) + {\rm Pi-D}_{{\rm normal}} \\
& = & -\left(P_{xx} \frac{\partial u_x}{\partial x} + P_{yy} \frac{\partial u_y}{\partial y} + P_{zz} \frac{\partial u_z}{\partial z}\right) \label{eq:pscompress}
\end{eqnarray}
\end{subequations}
so that
\begin{equation}
-({\bf P} \cdot \boldsymbol{\nabla}) \cdot {\bf u} = {\rm PDU} + {\rm Pi-D}_{{\rm shear}}.
\end{equation}
For an isotropic pressure with $P_{xx} = P_{yy} = P_{zz} \equiv P$, where $P$ is the scalar pressure, Eq.~(\ref{eq:pscompress}) reduces to ${\rm PDU} =  -P (\boldsymbol{\nabla} \cdot {\bf u})$, the known pressure dilatation from fluid mechanics. For an arbitrary pressure tensor, ${\rm PDU}$ gives the power density due to converging and diverging flows, which contains both dilatation and normal deformation.  Eq.~(\ref{eq:pscompress}) is the reasonable generalization of pressure dilatation when isotropy is not valid, as it contains contributions from dilatation in each direction independently.

\section{Physical Interpretation of Pressure-Strain Interaction}
\label{sec:physicalinterp}

Here, we provide the physical interpretation of the pressure-strain interaction contributions in the fluid and kinetic descriptions. The fluid description has partially been addressed previously \cite{del_sarto_pressure_2016,del_sarto_pressure_2018}. We provide simplified examples that allow for the physical interpretation to be made clear, with the idea that they can be used to motivate analogous processes for more general cases. While the fluid description is valid, both simulations and satellites now regularly measure the phase space density, measuring plasma properties at scales at and below the scales where treating a plasma as a fluid is no longer appropriate \cite{Shuster19,Shuster21}.  Thus, we argue it is important to develop a fully kinetic interpretation of the contributions to the pressure-strain interaction.  As shown in Sec.~\ref{sec:pidsign}, this understanding will provide insight into what it means to have a negative ${\rm Pi-D}$.

\subsection{Fluid Description of the Pressure-Strain Interaction}

\begin{figure}
\includegraphics[width=5.4in]{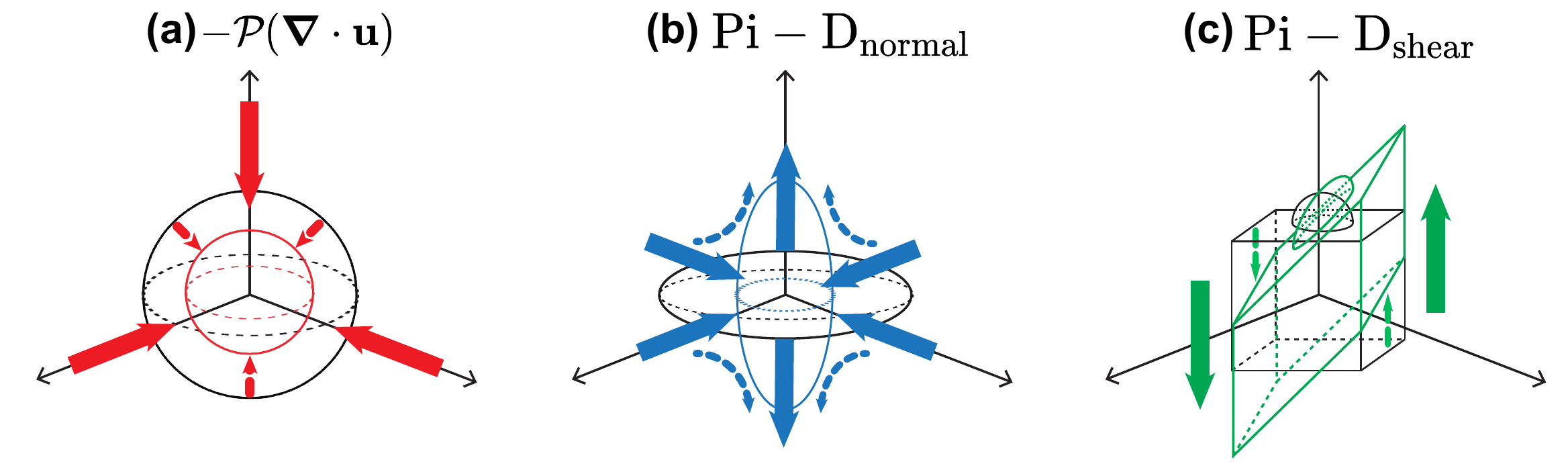}
\caption{\label{fig:dil_normdef_sheardef} Sketch of representative contributions to the pressure-strain interaction in the fluid description. Black shapes are the initial fluid elements, and bold arrows show the bulk flow directions. The dashed arrows map the change between initial and final shapes of the fluid elements. (a) Pressure dilatation (red), showing compression, (b) normal deformation (blue), and (c) shear deformation (green). Panels (a) and (b) are essentially copies of Figure 1 from ``Shear-induced pressure anisotropization and correlation with fluid vorticity in a low collisionality plasma,'' by Daniele Del Sarto and Francesco Pegoraro, Monthly Notices of the Royal Astronomical Society, {\bf 475}, 181 (2018) \cite{del_sarto_pressure_2018}; panel (c) is new.}
\end{figure}

We begin with the physical interpretation of pressure-strain interaction in the fluid description. It was treated in the limit of vanishing vector heat flux density ${\bf q}$ in Ref.~\cite{del_sarto_pressure_2018}, vividly conveyed in their Fig.~1 
that contains valid sketches of the effects of dilatation (red) and normal deformation (blue). However, because their analysis did not contain a vector heat flux density, the shear deformation term in Eq.~(\ref{eq:psincompress}) did not appear in their analysis. Thus, we extend their Fig.~1 in the general case in our Fig.~\ref{fig:dil_normdef_sheardef}. 

Panel (a) exemplifies pressure dilatation $-{\cal P} (\boldsymbol{\nabla} \cdot {\bf u})$, representing compression of the sketched spherical fluid element. Panel (b) exemplifies normal deformation ${\rm Pi-D}_{{\rm normal}}$, represented by the volume preserving change of shape of the sketched ellipsoidal fluid element. These two panels are modeled directly after Ref.~\cite{del_sarto_pressure_2018}. The initial fluid element is in black, the flow profile is in the large arrows, and the final fluid element is in color.  The small colored arrows denote the action of the fluid element due to the flow.

Panel (c) exemplifies shear deformation ${\rm Pi-D}_{{\rm shear}}$, which is not present in Ref.~\cite{del_sarto_pressure_2018}. The sheared flow deforms the fluid element, as in the standard treatment of flow shear in a fluid, except that this effect is purely collisionless. A key point is that shear deformation requires a non-zero off-diagonal pressure tensor element in the plane of the varying bulk flow and its gradient for there to be a contribution to the pressure-strain interaction [see Eq.~(\ref{eq:psincompress})]. Thus, we draw a cubical fluid element in (c) with a feature sticking out of the box to denote the need for the off-diagonal elements. Since the off-diagonal pressure tensor elements can be either positive or negative, shear deformation can lead to a positive or negative contribution to the pressure-strain interaction. Because the pressure-strain interaction is collisionless, any change in thermal energy due to it is formally reversible.  In contrast, collisional viscous heating is unable to lead to a decrease in thermal energy and is irreversible.

\subsection{Kinetic Description of ${\rm PDU}$}
\label{sec:kineticpdu}

Here we treat the kinetic theory interpretation of the pressure-strain interaction.  We do this by investigating how a phase space density evolves in time when there is a non-zero pressure-strain interaction to illustrate kinetically why there is a change in the thermal energy density.  We first emphasize that the pressure-strain interaction is local in space and time, and calculating it depends only on the local pressure tensor and the bulk flow velocity profile. Thus, instantaneously, determining if there is conversion between bulk flow and thermal energy density does not require knowledge of the presence of any body forces or collisions. 
In the treatment that follows, we ignore body forces and collisions. Although body forces and collisions are not needed to determine the local pressure-strain interaction, they do impact the motion of particles and the evolution of the phase space density, so these effects would have to be considered in addition to the phase space evolution considered here.
We briefly return at the end to motivate how body forces change the pictures that follow.

In the force-free, collisionless limit, the Boltzmann/Vlasov equation becomes
\begin{equation}
\frac{\partial f}{\partial t} + {\bf v} \cdot \boldsymbol{\nabla} f = 0. \label{eq:boltzconv}
\end{equation}
As is well known, this is merely a linear convection equation in position space at every ${\bf v}$.  We will use this in the examples that follow.

As an example which isolates ${\rm PDU}$, consider a plasma with a drifting bi-Maxwellian phase space density $f_{biM}$ aligned with a Cartesian coordinate system so that the pressure tensor ${\bf P}_{biM}$ is uniform in space and its elements are given by $P_{xx} = P_{yy} = P_\perp, P_{zz} = P_\|$, and $P_{jk} = 0$ for $j \neq k$. The effective pressure is then ${\cal P}_{biM} = (2 P_\perp + P_\|)/3$, and Eq.~(\ref{eq:pdecomppi}) reveals that the deviatoric pressure tensor $\boldsymbol{\Pi}_{biM}$ is 
\begin{equation}
  \boldsymbol{\Pi}_{biM} = {\bf P}_{biM} - {\cal P}_{biM} {\bf I} =  (P_\perp-P_\|) \left( \begin{array}{ccc}
    \frac{1}{3} & 0 & 0 \\ 0 & \frac{1}{3} & 0 \\ 0 & 0 & -\frac{2}{3}
    \end{array} \right).
\end{equation}
Using Eq.~(\ref{eq:ddef}), the associated ${\rm Pi-D}_{biM}$ for an arbitrary bulk flow profile ${\bf u}$ is
\begin{equation}
{\rm Pi-D}_{biM} = - \Pi_{jk} {\cal D}_{jk} = - (P_\perp - P_\|) \left[\frac{1}{3}
  (\boldsymbol{\nabla} \cdot {\bf u}) - \frac{\partial u_z}{\partial z} \right].
\end{equation}
As desired, this pressure tensor ${\bf P}_{biM}$ does not depend on flow shear even if it is present.
For definiteness, we consider $P_\| > P_\perp$.  We first treat converging flow in the parallel direction, such that ${\bf u} = u_z(z) {\bf \hat{z}}$, and for simplicity we treat bulk flow towards $z = 0$.

\begin{figure}
\includegraphics[width=5.4in]{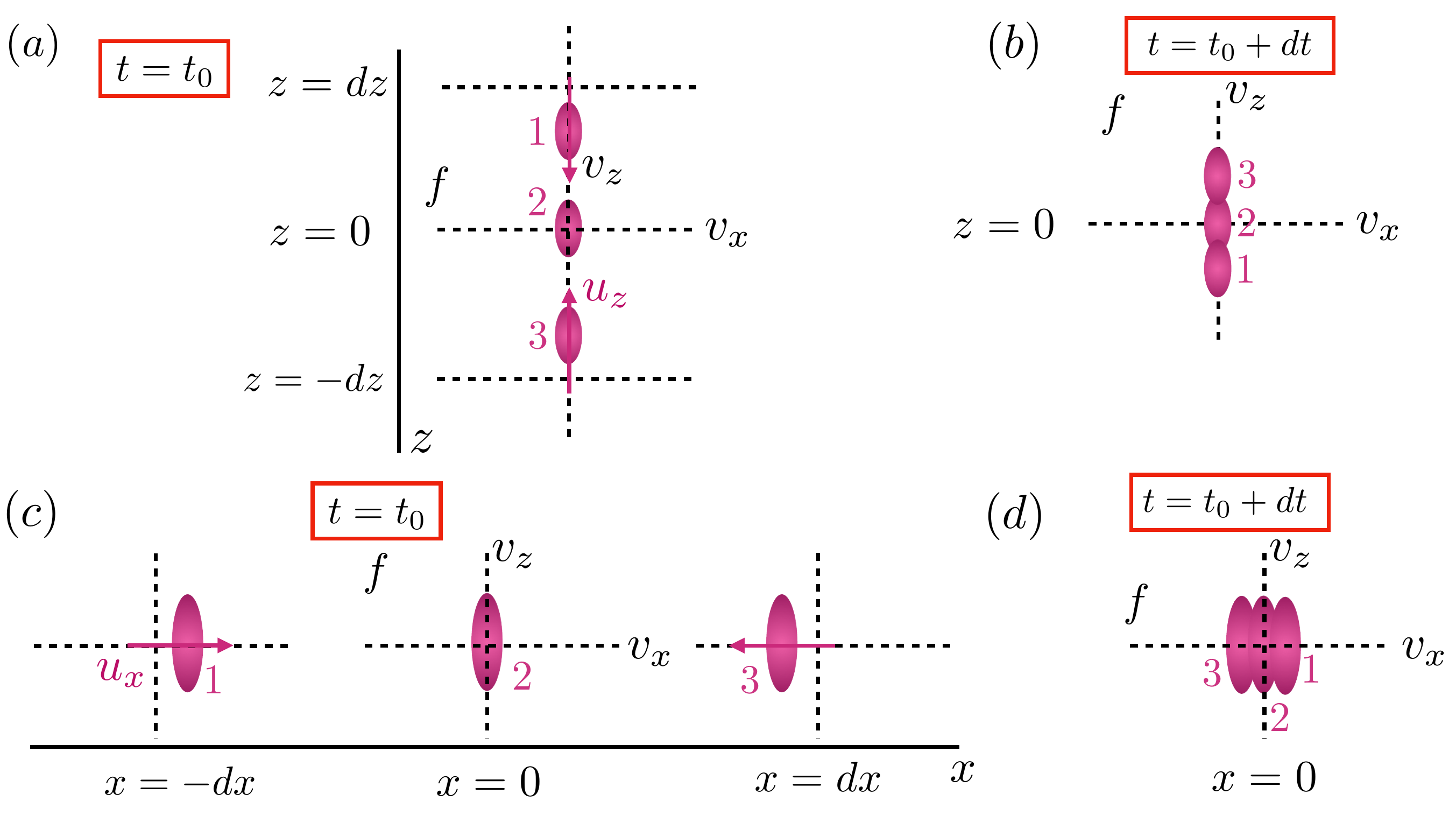}
\caption{\label{fig:convergebimax}  Sketches showing the physical interpretation of ${\rm PDU}, {\it i.e.,}$ heating via converging flow, in kinetic theory, ignoring body forces and collisions for simplicity.  Magenta ellipses denote a 2D slice of the phase space density $f$ in the $(v_x,v_z)$ plane given by bi-Maxwellian distributions with $P_\| > P_\perp$, where $x$ is a perpendicular direction and $z$ is parallel. (a) Phase space densities at initial time $t = t_0$ at three locations at and near $z = 0$.  The vertical bulk flow velocity $u_z$, denoted by the magenta arrows, is converging in the parallel direction.  (b) The phase space density at $z = 0$ at a slightly later time $t = t_0 + dt$. The phase space densities labeled 1, 2, and 3 in panel (a) evolve to their associated positions labeled in panel (b). The phase space density at this time is broader in $v_z$, implying an increase in thermal energy density.  Note, ${\rm Pi-D}$ is positive for this case.  (c) and (d) are analogous for the same phase space density except with converging bulk flow in $x$. There is an increase in the thermal energy density in the phase space density at $x=0$ at $t = t_0 + dt$ in panel (d).  Interestingly, ${\rm Pi-D}$ is negative for this case.}
\end{figure}

A sketch of the system at initial time $t = t_0$ is in Fig.~\ref{fig:convergebimax}(a). The phase space density $f_{biM}$ is sketched as the magenta ovals in the $(v_x,v_z)$ plane at three different spatial locations, $z = dz, 0,$ and $-dz$. The pressures are the same at each location, but the phase space densities are offset from the origin accordingly to impose that the bulk flow converges towards $z = 0$. A short time $dt$ later, $f_{biM}$ with $v_z < 0$ at $z > 0$ (labeled 1) convects down, $f_{biM}$ with $v_z > 0$ at $z < 0$ (labeled 3) convects up, and $f_{biM}$ near $v_z = 0$ at $z = 0$ (labeled 2) does not convect far, so the phase space density $f$ at $z = 0$ at $t = t_0 + dt$ is qualitatively displayed in Fig.~\ref{fig:convergebimax}(b), with the same numbering scheme to show where the particles came from at $t = t_0$. (We acknowledge that the precise phase space density at time $t_0 + dt$ would be affected by particles from cells beyond those plotted and would smear out the final distribution, but we do not attempt to capture this effect in the sketch for simplicity.) Comparing the phase space densities $f_{biM}$ and $f$ at $z = 0$ at $t = t_0$ and at $t = t_0 + dt$, respectively, we note that the breadth of $f$ in the perpendicular $v_x$ direction is the same as in $f_{biM}$ (there is no perpendicular heating), but $f$ is broader in the parallel $v_z$ direction than $f_{biM}$. Broadening a phase space density is the kinetic manifestation of heating, {\it i.e.,} increasing the thermal energy. This gives the kinetic interpretation of ${\rm PDU}$, {\it i.e.,} heating via converging flow in the $z$ direction, corresponding to $-P_{zz} (\partial u_z / \partial z)$ in Eq.~(\ref{eq:pscompress}).

We now consider converging flow in the perpendicular direction for the same initial phase space density, so now ${\bf u} = u_x(x) {\bf \hat{x}}$, treating bulk flow converging towards $x = 0$ for simplicity.  A sketch at the initial time $t = t_0$ is in Fig.~\ref{fig:convergebimax}(c), where the phase space density $f_{biM}$ is sketched at $x = -dx, 0,$ and $dx$.  Since $f_{biM}$ evolves in time according to the convection equation in Eq.~(\ref{eq:boltzconv}) in the absence of body forces and collisions, the phase space density $f$ at $x = 0$ a short time $dt$ later appears as sketched in Fig.~\ref{fig:convergebimax}(d). The phase space density $f$ does not broaden in the parallel $v_z$ direction, but does broaden in the $v_x$ direction. This is the kinetic manifestation of heating from ${\rm PDU}$ via converging flow in the $x$ direction, corresponding to $-P_{xx} (\partial u_x / \partial x)$ in Eq.~(\ref{eq:pscompress}). In both examples, heating due to converging flow contains contributions from both dilatation and normal deformation, a key point we return to in the next section.

Finally, we return to the effect of the presence of a body force ${\bf F}$. As stated earlier, it is clear from the expression for pressure-strain interaction that a body force cannot contribute to it, even though the forces impact the motion of the particles.  The sketches used here can still provide information for how to interpret the terms in the pressure-strain interaction when there is a body force present.  First consider a uniform body force, {\it i.e.,} it is the same at every position. The body force ${\bf F}$ changes the velocity of all particles of mass $m$ at a given position by the same increment $d{\bf v} = {\bf F} dt / m$ in a small increment in time $dt$, so it merely translates the phase space density in velocity space.  In the Lagrangian reference frame, this shift does not lead to a change in the thermal energy at the point in question beyond what is shown in the sketches in this section. If there is a force that is not uniform, a similar procedure happens except that the shift in velocity space of the particles is different at every location.  In our example, the phase space density is uniform in space, so the result is unchanged. In the more general case for which the phase space density is also not uniform, it would require a detailed analysis to understand the evolution of the particles and the associated phase space densities, which is beyond the scope of the present study.  However, we know the result for an arbitrary force and initial phase space density must be that the body force does not alter the pressure-strain interaction in the Lagrangian reference frame.

\subsection{Kinetic Description of ${\rm Pi-D}_{{\rm shear}}$}
\label{sec:kineticpidshear}

We next turn to the kinetic interpretation of heating via shear deformation.  As noted in the introduction \cite{Batchelor67,del_sarto_pressure_2016,del_sarto_pressure_2018}, the symmetric strain rate tensor ${\bf S}$ needs to be non-zero for the pressure-strain interaction to be non-zero, and the pressure-strain interaction is independent of the anti-symmetric strain rate tensor $\boldsymbol{\Omega}$.  Moreover, from Eq.~(\ref{eq:psincompress}), the pressure tensor must have a non-zero off-diagonal element in order for there to be heating via shear deformation.  Consequently, we consider flow shear of a phase space density with non-zero off-diagonal pressure tensor elements.

\begin{figure}
\includegraphics[width=5.4in]{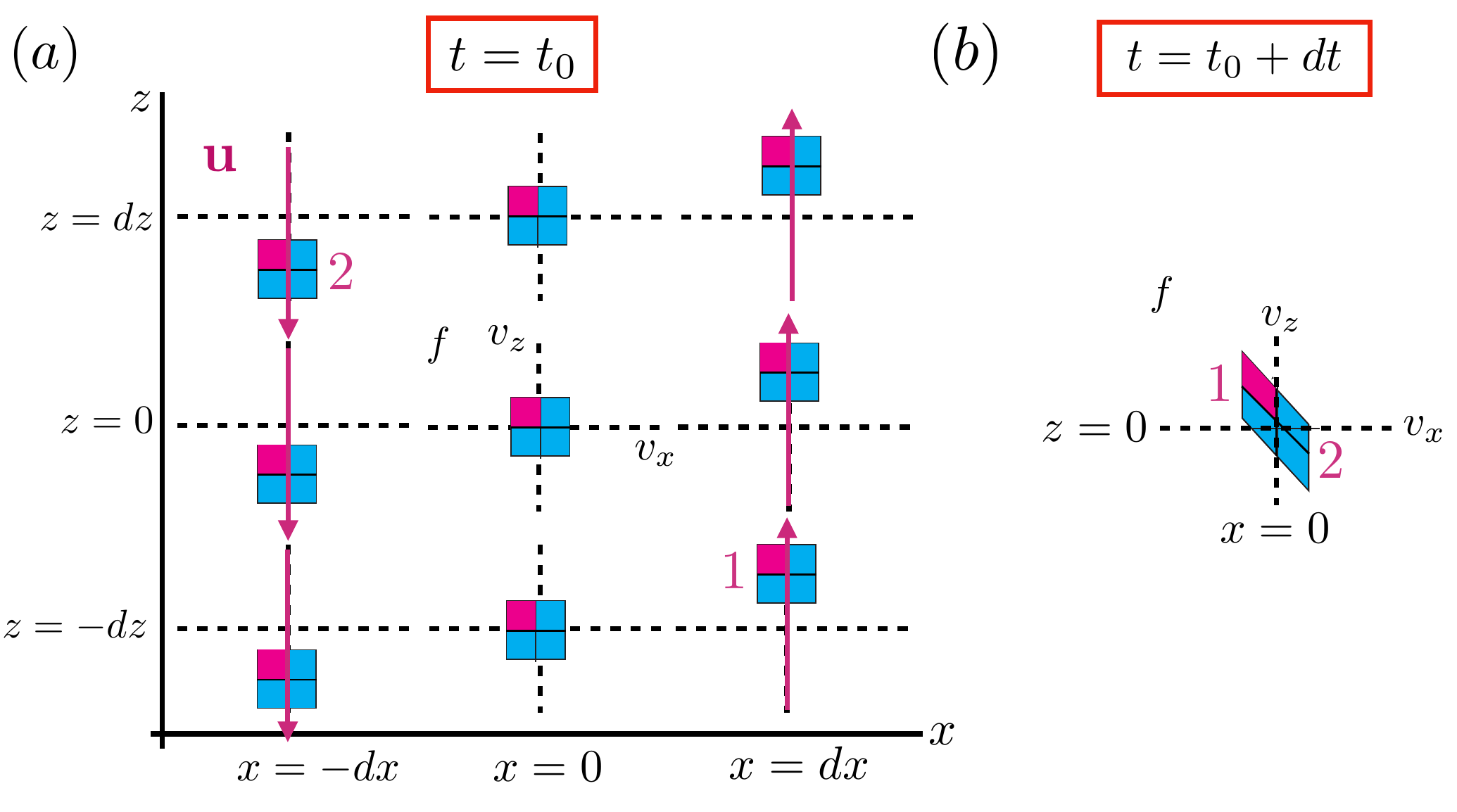}
\caption{\label{fig:shearheating} Sketch illustrating the kinetic theory explanation of why ${\rm Pi-D}_{{\rm shear}}$ leads to heating or cooling. (a) Array of sketches at locations in position space $(x,z)$ near the origin at the initial time $t_0$. Each sketch contains a phase space density $f$ in the $(v_x,v_z)$ plane in blue and red, where blue represents relatively low $f$ and red represents relatively high $f$. Such phase space densities have $P_{xz} < 0$. The placement of the phase space density in each axis system reveals its bulk flow ${\bf u}$, denoted for each $f$ by the magenta arrow. The flow profile has a representative form $(u_x,u_z) = (0, x)$.  (b) Sketch of the phase space density at the origin at a slightly later time $t = t_0 + dt$. The portions of the phase space densities in (a) labeled 1 and 2 evolve to make up the portions of the phase space densities in (b) labeled 1 and 2, respectively.  For this flow profile, there is a net displacement of particles away from the velocity space origin, implying an increase in thermal energy density.}
\end{figure}

Consider flow in the $xz$ plane; dynamics in the other planes is analogous. Figure~\ref{fig:shearheating}(a) contains sketches in a region near $(x,z) = 0$ at initial time $t = t_0$, with phase space densities sketched in $(v_x,v_z)$ space at an array of spatial locations given by $x = dx, 0,$ and $-dx$ and $z = dz, 0,$ and $-dz$.  In the kinetic picture, $P_{xz}$ is non-zero if the phase space density lacks symmetry in both the $v_x$ and $v_z$ directions relative to the bulk flow speed $(u_x,u_z)$. One way a phase space density can have a positive $P_{xz}$ is if it is elongated in the first and/or third quadrant in the $(v_x,v_z)$ plane compared to the second and/or fourth quadrants; similarly a negative $P_{xz}$ is elongated in the second and/or fourth quadrants in $(v_x,v_z)$ space. Another is if $f$ weighted higher in the first and/or third quadrants than the second and/or fourth quadrants. Figure~\ref{fig:shearheating}(a) includes a phase space density with $P_{xz} < 0$ due to the weighting of $f$ that is uniform in space, displayed with red signifying larger $f$ and blue signifying smaller $f$.  To impose a bulk velocity shear, the phase space densities are shifted relative to the velocity space origin, with the bulk flow direction denoted by the magenta arrows.  For this illustration, we assume a profile with ${\bf u} = u_z(x) {\bf \hat{z}}$, where $u_z$ is positive for $x > 0$ and negative for $u_z < 0$.

In the next increment in time $dt$, particles with $v_x > 0, v_z < 0$ at $(-dx,dz)$ (labeled 2) move towards the origin (in the absence of body forces and collisions), appearing in the $v_x > 0, v_z < 0$ quadrant at the origin at $t = t_0 + dt$ (labeled 2) in Fig.~\ref{fig:shearheating}(b). This portion of the phase space density is blue, meaning $f$ is relatively low there. Similarly, in the phase space density at $(dx,-dz)$, particles in the left part of the distribution (labeled 1) have $v_x < 0, v_z > 0$, so they also move toward the origin.  At $t = t_0 + dt$, they become the population in the $v_x < 0, v_z > 0$ portion of the phase space density (labeled 1) in Fig.~\ref{fig:shearheating}(b). The red portion in Fig.~\ref{fig:shearheating}(b) is the portion of the phase space density at $t = t_0 + dt$ with higher $f$ values than elsewhere in the phase space density.

To interpret this result, we note the phase space density at the origin at $t = t_0 + dt$ effectively stretches away from the velocity space origin in the second and fourth quadrants relative to the phase space density at the origin at $t = t_0$, and moves closer to the velocity space origin in the first and third quadrants than at $t = t_0$.  If $f$ had begun at $t = t_0$ as symmetric in $(v_x,v_z)$ space ({\it i.e.,} if $P_{xz}$ had been 0), this would lead to no net heating at $t = t_0 + dt$, since there would be equal numbers of particles brought closer to the velocity space origin as those brought further away.  However, in this case, there are more particles in quadrant 2 than the other quadrants at $t = t_0 + dt$, so there are more particles further from the origin.  This is the kinetic manifestation of heating. This example provides motivation for the kinetic theory of heating via pressure-strain interaction due to ${\rm Pi-D}_{{\rm shear}}$, with the same caveat as in the previous subsection that body forces and collisions can alter the particle trajectories and phase space density evolution, but cannot directly impact the pressure-strain interaction. We note that $P_{xz} < 0$ and $\partial u_z / \partial x > 0$ in this example, so the term $-P_{xz}(\partial u_z/\partial x)$ in Eq.~(\ref{eq:psincompress}) is positive. This is associated with heating, consistent with the physical picture given here.  A similar construction with the higher $f$ region in the fourth quadrant (so that $P_{xz}$ is again negative) also leads to heating.  If the higher $f$ region is in the first or third quadrant, it would lead to cooling because there are more particles closer to the velocity space origin than farther away from it.  In this case $P_{xz}$ is positive, and the ${\rm Pi-D}_{{\rm shear}}$ contribution to the pressure-strain interaction is negative, consistent with cooling.  Thus, 
${\rm Pi-D}_{{\rm shear}}$ can contribute to heating or cooling depending on the flow profile and the sign of the off-diagonal pressure-tensor elements, and it is in principle reversible.

\begin{figure}
\includegraphics[width=5.4in]{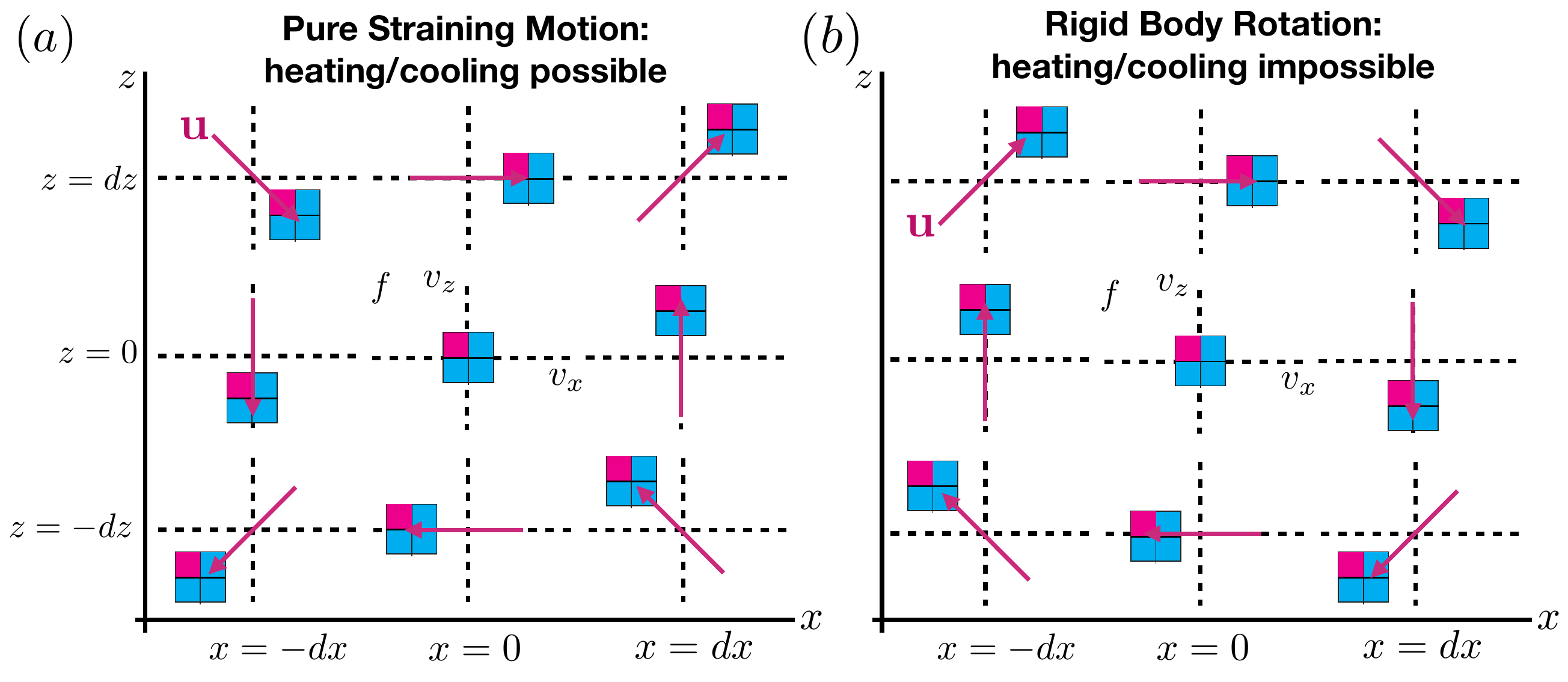}
\caption{\label{fig:shearvortical} Sketch illustrating the kinetic theory explanation of why (a) symmetric shear (pure straining motion) can lead to heating/cooling, while (b) anti-symmetric shear (rigid body rotation) cannot.  In each sketch, the grid in $x$ and $z$ denotes physical positions in the environment of the origin $(x,z) = (0,0)$.  Each red and blue box denotes a phase space density $f$ with a negative $P_{xz}$ in the $(v_x,v_z)$ plane at the location in question. The local bulk flow ${\bf u}$ is denoted for each $f$ with a magenta arrow.  The flow profiles are (a) $(u_x,u_z) = (z,x)$ and (b) $(u_x,u_z) = (z, -x)$.}
\end{figure}

We conclude this subsection with a kinetic theory interpretation of why pure straining motion leads to a contribution to the pressure-strain interaction, but rigid body rotation does not.  Figure~\ref{fig:shearvortical}(a) is a sketch analogous to Fig.~\ref{fig:shearheating}(a) of a hyperbolic bulk flow profile corresponding to pure straining motion with $\partial u_k / \partial r_j = \partial u_j / \partial r_k$, so $\boldsymbol{\Omega} = 0$. Analogous to Fig.~\ref{fig:shearheating}(b), there is a flow of particles towards (and away from) the origin in the next small increment in time, which serves to increase the thermal energy density at the origin at $t = t_0 + dt$.  In contrast, Fig.~\ref{fig:shearvortical}(b) shows a similar sketch, but for rigid body rotation for which $\partial u_k / \partial r_j = -\partial u_j / \partial r_k$, so ${\bf S} = 0$.  In this case, the flow profile imposes that the particles in the phase space densities surrounding the origin predominantly go around the origin rather than changing the phase space density at the origin, leaving the thermal energy density at the origin unchanged.  This is the kinetic explanation for why rigid body rotation does not contribute to the pressure-strain interaction.

\section{Kinetic Interpretation of Normal Deformation and Implications for the Sign of ${\rm Pi-D}$}
\label{sec:pidsign}

It has been shown numerically and observationally that ${\rm Pi-D}$, {\it i.e.,} collisionless viscous heating, can be positive or negative, which has been puzzling because collisional viscous heating must be non-negative. We use the present results to interpret the sign of ${\rm Pi-D}$ within the kinetic description. We emphasized in Sec.~\ref{sec:piddecomposition} that ${\rm Pi-D}$ contains both normal deformation and shear deformation. Consequently, it is not a well-posed question to ask what a negative value of ${\rm Pi-D}$ means physically because it is ambiguous; it could have contributions from either term.  The physics of ${\rm Pi-D}_{{\rm shear}}$ was discussed in Sec.~\ref{sec:kineticpidshear}, including what it means physically for it to be positive or negative. The example of heating due to converging flow in Sec.~\ref{sec:kineticpdu} explained the sign only for ${\rm PDU}$, {\it i.e.}, the sum of the dilatation and normal deformation terms, so we reconsider the example given there to isolate the normal deformation and explain the kinetic interpretation of its sign.

Consider again the bi-Maxwellian distribution $f_{biM}$ with $P_{xx} = P_{yy} = P_\perp$ and $P_{zz} = P_\|$ with $P_\| > P_\perp$ discussed in the previous section.  For the two bulk flow profiles in Fig.~\ref{fig:convergebimax}, analytic expressions for ${\rm Pi-D}_{{\rm normal}}$ are readily calculated from Eq.~(\ref{eq:piddeform}). For parallel converging flow ${\bf u} = u_z(z) {\bf \hat{z}}$, we get
\begin{equation}
  \boldsymbol{{\cal D}}_{{\rm normal}} = \boldsymbol{{\cal D}} = \frac{\partial u_z}{\partial z} \left( \begin{array}{ccc}
    -\frac{1}{3} & 0 & 0 \\ 0 & -\frac{1}{3} & 0 \\ 0 & 0 &
    \frac{2}{3} \end{array} \right),
\end{equation}
so that
\begin{equation}
{\rm Pi-D}_{{\rm normal}} = {\rm Pi-D} = \frac{2}{3} \frac{\partial u_z}{\partial z} (P_\perp - P_\|). \label{eq:pidconvergingz}
\end{equation}
A similar derivation reveals that if ${\bf u} = u_x(x) {\bf \hat{x}}$ for perpendicular converging flow,
\begin{equation}
{\rm Pi-D}_{{\rm normal}} = {\rm Pi-D} = - \frac{1}{3} \frac{\partial u_x}{\partial x} (P_\perp - P_\|). \label{eq:pidconvergingx}
\end{equation}
Importantly, ${\rm Pi-D}_{{\rm normal}}$ for parallel and perpendicular converging flow have opposite signs.

For the flow profiles in Fig.~\ref{fig:convergebimax}, ${\rm Pi-D}_{{\rm normal}}$ is positive for parallel converging flow but negative for perpendicular converging flow.  We know the net dilatation plus normal deformation due to converging flow leads to heating as quantified by ${\rm PDU}$ for converging flow in either direction.  Thus, it may not be surprising that ${\rm Pi-D}_{{\rm normal}} > 0$ for parallel converging flow. However, it is counterintuitive that ${\rm Pi-D}_{{\rm normal}} < 0$ for perpendicular converging flow because negative ${\rm Pi-D}$ has been referred to as ``cooling.''

The resolution of this apparent paradox is to consider ${\rm Pi-D}_{{\rm normal}}$ in the context of pressure dilatation $-{\cal P} (\boldsymbol{\nabla} \cdot {\bf u})$.  For the perpendicular converging flow case,
\begin{equation}
  -{\cal P} (\boldsymbol{\nabla} \cdot {\bf u}) = - {\cal P} \frac{\partial
    u_x}{\partial x} = - \left(\frac{2P_\perp + P_\|}{3}\right)
  \frac{\partial u_x}{\partial x}. \label{eq:pressdilx}
\end{equation}
This quantity is non-negative for converging flow, which reflects that there is heating.  The sum of dilatation and normal deformation from Eqs.~(\ref{eq:pidconvergingx}) and (\ref{eq:pressdilx}) to get ${\rm PDU}$ is
\begin{equation}
  {\rm PDU} = - P_\perp \frac{\partial u_x}{\partial x}, \label{eq:pdivuconverging}
\end{equation}
as expected from Eq.~(\ref{eq:pscompress}). This is positive for converging flow, corresponding to a net heating, as expected.

This simple example suggests the kinetic theory interpretation of ${\rm Pi-D}_{{\rm normal}}$.  The quantity ${\rm PDU}$ describes the total volumetric heating rate due to converging flow.  Pressure dilatation describes the volumetric rate of compressible heating if the system was in equilibrium with a (scalar) pressure ${\cal P}$ since it has the form $-{\cal P}(\boldsymbol{\nabla} \cdot {\bf u})$. Kinetically, pressure dilatation describes the heating that would take place if the phase space density $f$ were replaced by a Maxwellian distribution $f_M$ with the density $n = \int f d^3v$ and with its pressure $P$ given by the effective pressure ${\cal P}$ found from the local phase space density $f$. The phase space density $f_M$ is known as the Maxwellianized distribution of $f$. The contribution from ${\rm Pi-D}_{{\rm normal}}$, then, describes the correction to the total of the volumetric heating rate due to converging flow due to the phase space density $f$ not being the Maxwellianized distribution $f_M$.

In the example of the bi-Maxwellian distributions with $P_\| > P_\perp$ in Fig.~\ref{fig:convergebimax}(a), the Maxwellianized distribution $f_M$ of the phase space density $f_{biM}$ is round in velocity space, so it is cooler in the parallel direction and hotter in the perpendicular direction than $f$.  Compression of $f_M$ in the perpendicular $x$ direction would heat the plasma, making the phase space density at $t = t_0 + dt$ broader in the $v_x$ direction than it would be due to compression of $f_{biM}$. This means there would be a higher volumetric heating rate of $f_M$ than there would be of $f_{biM}$ since $P_\| > P_\perp$. Therefore, ${\rm Pi-D}_{{\rm normal}}$ is negative in this example because it represents the correction to the volumetric heating rate of the actual phase space density $f_{biM}$ because it is not the Maxwellianized distribution $f_M$; in this case, the correction is negative.  Thus, ${\rm Pi-D}_{{\rm normal}} < 0$ during converging flow does not represent physical cooling, because when combined with $-{\cal P} (\boldsymbol{\nabla} \cdot {\bf u})$ to form ${\rm PDU}$, the volumetric heating rate into thermal energy is positive, as is expected for converging flow.  This is in contrast to ${\rm Pi-D}_{{\rm shear}}$, which necessarily contributes to cooling when it is negative.

Similar reasoning holds for parallel converging flow.  As shown in Fig.~\ref{fig:convergebimax}(b), the result of converging parallel flow is to generate a phase space density at the origin that is even more elongated than $f_{biM}$ at $t = 0$.  In this case, pressure dilatation is again positive because there is converging flow, but here ${\rm Pi-D}_{{\rm normal}}$ is also positive.  This is because the Maxwellianized distribution $f_M$ is narrower in the parallel direction than $f_{biM}$, so the heating of $f_M$ is less than the heating of $f_{biM}$.  The contribution from ${\rm Pi-D}_{{\rm normal}}$ is positive to make up for the part of the heating omitted from the converging flow acting on $f_M$.

Thus, simply knowing that ${\rm Pi-D}$ is negative is insufficient to know if it is caused by normal deformation or shear deformation, and it is insufficient to know if there is overall cooling via the pressure-strain interaction.  If ${\rm Pi-D} < 0$ and ${\rm Pi-D}_{{\rm shear}}$ dominates, there is a contribution towards cooling. However, if ${\rm Pi-D} < 0$ and is dominated by ${\rm Pi-D}_{{\rm normal}}$, one cannot know if there is heating or cooling due to converging or diverging flow because $-{\cal P} (\boldsymbol{\nabla} \cdot {\bf u})$ can have either sign depending on whether there is converging or diverging flow.  It is ${\rm PDU}$ that must be measured to assess if heating/cooling due to converging/diverging flow is taking place.

\section{Discussion and Conclusions}
\label{sec:discuss}

The pressure-strain interaction, including ${\rm Pi-D}$, has undergone intense scrutiny in the past few years because it concisely describes the rate that energy density is converted between bulk flow and thermal. Pressure dilatation is the portion of pressure-strain interaction associated with compression and expansion, while ${\rm Pi-D}$ is the portion associated with incompressible heating \cite{del_sarto_pressure_2018} and has been described as collisionless viscosity \cite{yang_PRE_2017}.  Despite the scrutiny, fundamental questions about the physical interpretation of the pressure-strain interaction and ${\rm Pi-D}$ have persisted, including what it means for ${\rm Pi-D}$ to be negative. 

In this study, we use the fact that ${\rm Pi-D}$ contains both normal deformation and shear deformation to propose an alternate decomposition of the pressure-strain interaction with the ${\rm PDU}$ and ${\rm Pi-D}_{{\rm shear}}$ terms, which separate the pressure-strain interaction into the power densities associated with converging/diverging flow and flow shear, respectively.  The ${\rm PDU}$ term is a combination of the dilatation and normal deformation terms, and gives the reasonable generalization of dilatation for systems not in local thermodynamic equilibrium.  In the large magnetic field limit, it was shown \cite{Hazeltine13} that the ${\rm Pi-D}_{{\rm shear}}$ term ($-\boldsymbol{\pi}_{gv} : U$ in their notation) vanishes to low order in the strong magnetic field expansion. This is because the magnetic field dominates all other collisionless physics in the limit in question, and the magnetic field itself does not directly contribute to the pressure-strain interaction [see Eq.~(\ref{eq:ethermaleqn}) and Ref.~\cite{del_sarto_pressure_2016}]. Outside of this limit, as shown here, ${\rm Pi-D}_{{\rm shear}}$ need not vanish.

Using these results, we provide a physical understanding of the contributions to the pressure-strain interaction both from a fluid perspective, with one modification from previous work on the subject that ignored the vector heat flux density \cite{del_sarto_pressure_2018}, and fully in the kinetic picture at the phase space density-level.  We use the results to explain kinetically why pure straining motion (a symmetric strain rate tensor) can lead to a change in thermal energy but rigid body rotation (an asymmetric strain rate tensor) cannot.  We finally use these results to give the physical mechanisms that cause ${\rm Pi-D}$, including giving a new kinetic theory interpretation for the normal deformation term.  We further show the counterintuitive result that while converging flow must contribute to a positive pressure-strain interaction, it can contribute to a negative ${\rm Pi-D}$ for systems not in LTE. 

We emphasize a number of consequences of this study that may be of use to the field:
\begin{enumerate}
\item As has been recognized elsewhere \cite{Yang17,Matthaeus20,Yang_2022_ApJ}, the pressure-strain interaction $-({\bf P} \cdot \boldsymbol{\nabla}) \cdot {\bf u}$ is the most relevant quantity to determine the rate of change of bulk flow energy density into thermal energy density (heating or cooling), rather than ${\rm Pi-D}$ in isolation.
\item It is correct that the pressure dilatation $-{\cal P} (\boldsymbol{\nabla} \cdot {\bf u})$ describes the volumetric rate of heating/cooling due to compression/expansion, but it is  not the full description of energy conversion in converging or diverging flow. Similarly, ${\rm Pi-D}$ is the measure of incompressible heating, but contains both normal deformation and shear deformation.  In contrast, ${\rm PDU}$ gives the effect of converging/diverging flows, and ${\rm Pi-D}_{{\rm shear}}$ gives the effect of flow shear.  We believe both decompositions have merit for analyzing the energy conversion in physical processes and provide complementary information. We envision that keeping all three terms -- pressure dilatation, ${\rm Pi-D}_{{\rm normal}}$, and ${\rm Pi-D}_{{\rm shear}}$ -- may also prove useful in some circumstances.
\item A local measurement of a negative ${\rm Pi-D}$ does not imply there is cooling. If ${\rm Pi-D}$ is negative due to normal deformation, the net effect of normal deformation and dilatation in the total pressure-strain interaction is still positive if the flow is converging.  Meanwhile, a negative ${\rm Pi-D}$ could also be the result of shear deformation, so there is no way to unambiguously identify the key physical processes at play from the sign of ${\rm Pi-D}$ alone.
\item The physical interpretation of the normal deformation portion of ${\rm Pi-D}$ in kinetic theory is the difference between the rate of compressional heating and the rate of compressional heating of the same process were the phase space density replaced by a Maxwellian distribution of the same effective pressure.
\item The introduction of the traceless strain-rate tensor $\boldsymbol{{\cal D}}$, which has been carried into plasma physics following a long history in the study of neutral fluids \cite{Batchelor67}, is only advantageous to study the rate of heating/cooling that is compressible vs.~incompressible.  However, it is not useful for distinguishing the heating between converging/diverging flows and flow shear.  While the difference may be negligible in neutral fluids that are near local thermodynamic equilibrium, they can be very different in plasmas that are far from local thermodynamic equilibrium.
\item It bears noting that the pressure-strain interaction is rigorously the quantity that describes the rate of conversion between bulk flow and thermal energy density, but it is not the only term that determines the local thermal energy density. In particular, thermal energy density flux and/or heat flux can also change the local thermal energy density \cite{Du20,song_forcebalance_2020}, even though these terms do not contribute to changes in the net thermal energy in a closed or isolated system \cite{Yang17}.  
\item The thermal energy density describes the random energy in a phase space density, {\it i.e.,} $\int (1/2) m v^{\prime 2} f d^3v$. However, other forms of energy such as $\int (1/2) m v_x^{\prime} v_y^\prime f d^3v$ or higher order moments are not contained in the thermal energy density, yet represent a possible energy channel during a physical process.  The energy going into channels beyond thermal energy density is treated in a separate study \cite{Cassak_FirstLaw_2022}.
\end{enumerate}

In Paper II, we derive the pressure-strain interaction in magnetic field-aligned coordinates.  In Paper III, we display the pressure-strain interaction in Cartesian and magnetic field-aligned coordinates in PIC simulations and use the results to determine the mechanisms that contribute to the pressure-strain interaction during collisionless reconnection.  For future work, it would be interesting to employ the decomposition of the pressure-strain interaction discussed here more broadly in simulation data and observational data to separate converging/diverging flow effects from shear flow effects. Example systems where such studies would be interesting include collisionless reconnection, plasma turbulence, and collisionless shocks. 

Also, we again point out from Eqs.~(\ref{eq:ethermaleqn}) and (\ref{eq:ekineticeqn}), the presence of collisions and body forces such as electric, magnetic, or gravitational forces, enters directly into the bulk flow energy density equation but not directly into the pressure-strain interaction. Body forces are quantified by the $n{\bf u} \cdot {\bf F}$ term in Eq.~(\ref{eq:ekineticeqn}), which for the electromagnetic force is $q n {\bf u} \cdot {\bf E}$ for a species of charge $q$.  This quantity, including the version summed over species given by ${\bf J} \cdot {\bf E}$, has also been under intense scrutiny in the study for describing the conversion between bulk flow energy and electromagnetic energy \cite{Zenitani11,Klein16,Burch16b,Wilder18,Chen19,Afshari21}. A better understanding of how body forces impact thermal energy should remain a topic of future work \cite{Howes17}.

\begin{acknowledgments}
We acknowledge beneficial conversations with Yan Yang.  We gratefully acknowledge support from NSF Grant PHY-1804428, DOE grant DE-SC0020294, and NASA grant 80NSSC19M0146.  This research uses resources of the National Energy Research Scientific Computing Center (NERSC), a DOE Office of Science User Facility supported by the Office of Science of the US Department of Energy under Contract no.~DE-AC02-05CH11231.
\end{acknowledgments}

\providecommand{\noopsort}[1]{}\providecommand{\singleletter}[1]{#1}%

\end{document}